\documentclass{article}
\usepackage{spconf,amsmath,graphicx}

\usepackage{graphicx}
\usepackage{verbatim}
\usepackage{mwe}
\usepackage[nospace]{cite}

\usepackage{amssymb}
\usepackage{amsfonts}
\usepackage{bbm}
\usepackage{hyperref}
\usepackage{color}
\usepackage{float}

\usepackage{array}
\usepackage{multirow}
\usepackage{subcaption}

\usepackage{algorithmic}
\usepackage{algorithm2e}
\usepackage{algorithm2e,setspace}

\title{STEP-GAN: A Step-by-Step Training for Multi Generator GANs with application to Cyber Security in Power Systems}
\name{Mohammad Adiban$^\dagger$, Arash Safari$^*$, Giampiero Salvi$^\dagger$ \thanks{G. Salvi is  also  affiliated  with  KTH  Royal  Institute  of  Technology,  School  of Electrical Engineering and Computer Science, Stockholm, Sweden.}
}
\address{$^\dagger$Electronic Systems Department, Norwegian University of Science and Technology, NO-7491 Norway\\$^*$School of Electrical and Computer Engineering, University of Tehran, Tehran, Iran
	\\E-mails: \{mohammad.adiban,giampiero.salvi\}@ntnu.no, \& safari.arash@ut.ac.ir}
\begin{document}
\ninept
\maketitle

\begin{abstract}
In this study, we introduce a novel unsupervised countermeasure for smart grid power systems, based on generative adversarial networks (GANs). Given the pivotal role of smart grid systems (SGSs) in urban life, their security is of particular importance. In recent years, however, advances in the field of machine learning, have raised concerns about cyber attacks on these systems. Power systems, among the most important components of urban infrastructure, have, for example, been widely attacked by adversaries. Attackers disrupt power systems using false data injection attacks (FDIA), resulting in a breach of availability, integrity, or confidential principles of the system. Our model simulates possible attacks on power systems using multiple generators in a step-by-step interaction with a discriminator in the training phase. As a consequence, our system is robust to unseen attacks. Moreover, the proposed model considerably reduces the well-known mode collapse problem of GAN-based models. Our method is general and it can be  potentially employed in a wide range of one of one-class classification tasks. The proposed model has low computational complexity and outperforms baseline systems about 14\% and 41\% in terms of accuracy on the highly imbalanced publicly available industrial control system (ICS) cyber attack power system dataset.
\end{abstract}


\section{Introduction}
\label{sec:introduction}
Cyber-physical systems (CPS) are defined as transformative technologies that integrate interconnected systems, including physical devices and computational networks in order to control and monitor physical processes~\cite{lee2015cyber}. A Smart Grid System (SGS), as one of the primary types of CPS, consists of a variety of operational and energy measurements, including smart meters and appliances, renewable energy resources, and energy efficiency resources~\cite{yadav2016review}. Due to the development and importance of smart grid systems, concerns about the security of SGSs are significantly increasing, inasmuch as these systems are subjectable to cyber attacks and since reports of sabotage and spoofing attacks on smart systems are increasing~\cite{adiban2017sut, adiban2020replay}. Furthermore, with the advances in technology, cyber-attacks are becoming more blended and sophisticated, enabling adversaries to easily target multiple layers of a power system simultaneously~\cite{langner2011stuxnet, bencsath2012duqu, el2018cyber}. These attacks aim to affect the performance of the system by altering the transmitted data or adding manipulated data that leads to a breach of availability, integrity, or confidential principles~\cite{gunduz2018analysis}.

Among smart grids, power systems are one of the most effective and widespread infrastructure components on which today's society is increasingly dependent.
These grids are significantly threatened by cyber attacks~\cite{teixeira2010cyber, karimipour2017robust, lai2019robustness, chen2020study}. To assure the safety and reliability of operations in power systems, many crucial data measurements, including electricity power flow and bus power injection, are continuously monitored~\cite{liu2014detecting}. In the next step, the monitored data is transmitted to a dispatching center system by a state estimation function of a supervisory control and data acquisition (SCADA) system that obtains the real-time operation state of power systems~\cite{ruan2017interval}. Most of the studies that attempt to detect cyber attacks in smart grid power systems automatically, use some form of supervision assuming the nature of the attacks is known~\cite{sakhnini2019smart, zhang2017deep, yan2016detection, jahromi2019deep, hink2014machine}. While collecting attack data is a challenging process, with the advancement of technology, attackers are able to use a variety of sophisticated methods to perform innovative cyber attacks, making it difficult to predict the nature of the attacks. Furthermore, labeled data is not always available in the real world, and data labeling is a costly and time-consuming process, requiring the use of human resources that can be associated with human error.

An alternative approach is to treat attack detection in smart grid systems as an unsupervised machine learning task. However, many of the unsupervised techniques related to anomaly detection in smart grid systems have been developed based on linear projection and transformation that is inadequate to deal with the inherent non-linearity of multivariate time series data (e.g power systems data)~\cite{ghosh2018multi}. Also, most current methods use a simple comparison between the current mode and the predicted normal range to detect cyber attacks, which is mostly insufficient owing to the inherent complexity and high dimensionality of smart grids data. Recently, the generative adversarial networks (GANs) have been proposed to anomaly detection tasks~\cite{zheng2019one, zenati2018efficient, zenati2018adversarially}. Although GAN models have been widely and successfully used in many tasks such as image processing, their use in time-series data processing is much more rare despite their successful performance in generating time-series sequences~\cite{ mogren2016c, esteban2017real, ghosh2018multi}. One of the main problems that GANs suffer from is mode collapse~\cite{srivastava2017veegan} which refers to the problem of missing some of the modes of the multi-modal data it was trained on.

\begin{figure}
\centering
\includegraphics[width=0.32\textwidth]{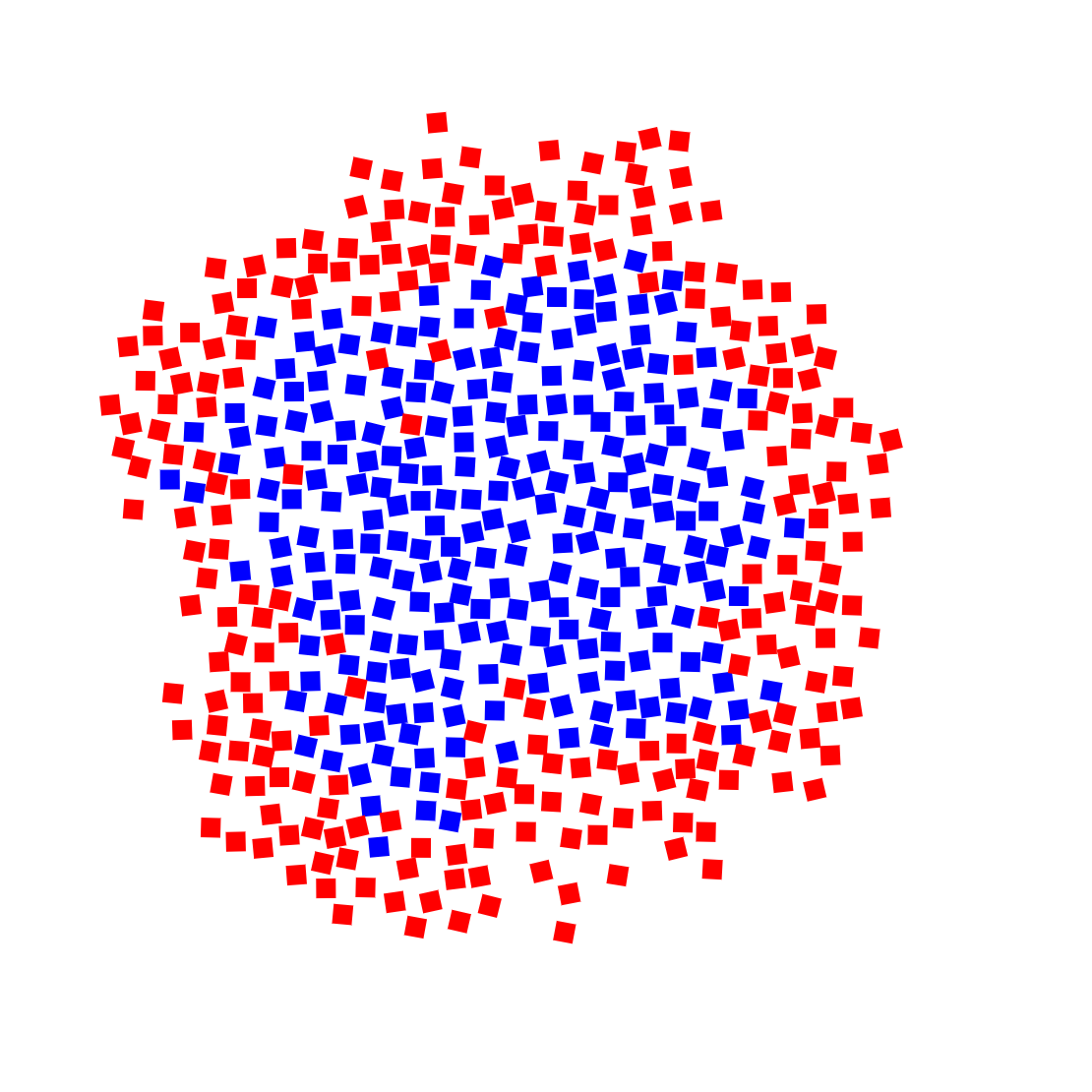}
        \caption{A simple schematic of complementary distribution. Blue points indicate normal data and red points follow the complementary distribution of normal data.}
        \label{fig:normal_attack_data}
\end{figure}

In this study, we propose a novel unsupervised GAN-based countermeasure to detect anomalies on smart grid power systems, which is inspired by~\cite{ ghosh2018multi} and \cite{zheng2019one}. Unlike many traditional techniques, our proposed method detects anomalies in an unsupervised fashion regardless of the nature of the attacks. The proposed model attempts to simulate the distribution of attack data, assuming that logically possible attacks on the power system have a complementary distribution~\cite{zheng2019one, dai2017good} close to the real data distribution. Here, complementary distribution refers to a term in which the generated data define a distribution that approaches the normal data distribution with minimal overlap (see Figure~\ref{fig:normal_attack_data}). Intuitively, attack data capable of deceiving power systems is expected to follow the complementary distribution of normal data.

More importantly, our approach provides a new technique for resolving mode collapse problems and seeks to model potential modes in which attacks (anomalies) can occur. As a result, our system is potentially more robust to unseen attacks. Our contributions are summarized as follow:
\begin{itemize}
    \item we propose a novel step-by-step training for multi generators GAN-based countermeasure against False Data Injection (FDI) attacks on power systems for the first time, which is robust to unseen attacks,
    \item we show that our model considerably reduces the well-known \textit{mode collapse} issue of GANs,
    \item the proposed method is general and it can be used in a wide range of tasks,
    \item we demonstrate that the proposed method significantly outperforms the baseline systems in terms of accuracy and F-measure on the highly imbalanced publicly available industrial control system (ICS) cyber attack power system dataset.
\end{itemize}

The remainder of this paper is organized as follows. First, we list the related studies in Section \ref{sec:related-works}. Then, the proposed method is introduced in Section \ref{sec:proposed-system}. Subsequently, we give a brief explanation of the experimental setup in Section \ref{sec:Experimental-setup}. Results and discussion are presented in Section \ref{sec:results} and \ref{sec:discussion}. Finally, we conclude the paper in Section \ref{sec:conclusion}.\\

\section{Related Works}
\label{sec:related-works}
Many cyber physical attack detection techniques have been developed in recent years. In this section, we present a brief discussion of the methods in this area. Lee et al. ~\cite{lee2014cyber} proposed a method to model the malicious cyber attack inside the system estimation using the expectation-maximization (EM) algorithm. They used the EM algorithm to find missing data and to optimize intractable likelihood function. The work presented in~\cite{pan2015classification} provided a sequential pattern mining approach to accurately extract patterns of power-system disturbances and cyber-attacks from heterogeneous time-synchronized data, consisting of synchrophasor measurements, relay logs, and network event monitor logs. Karimipour et al.~\cite{karimipour2017robust} presented a state estimation algorithm to detect FDI attacks in power systems. To this aim, they proposed an analytical technique based on the Markov chain theory and Euclidean distance metric. Their results show improvement in the traditional bad data detection method. In another work~\cite{jahromi2019deep}, the authors proposed an unsupervised neural network, called autoencoders (AE)~\cite{ng2011sparse} to extract meaningful features from power systems data. Then, they employed multiple traditional machine learning based classifiers, including artificial neural network, decision Tree, K-nearest neighbor, random forest, gradient boosting, and Adaboost in order to detect cyber-attacks. Basumallik et al.~\cite{basumallik2019packet} proposed a convolutional neural network (CNN) data filter with Nesterov Adam gradient descent and categorical cross-entropy loss to validate the phasor measurement units (PMU) data. This filter extracts inter time-series relationships to classify different power system events by comparing the temporal structure of PMU packet data.~\cite{chhetri2019gan} proposed a security model based on the Conditional Generative Adversarial Network (CGAN)~\cite{mirza2014conditional} which models the conditional probability distribution between the various information flow that provides a theoretical foundation to enable a system-level methodology for the design and analysis of cyber physical production systems against cross-domain attacks. They also used a generation algorithm to search and prune the graph in order to reduce the complexity of the model. Hassan et al.~\cite{hassan2020increasing} proposed a supervised method using a combination of the Random Subspace (RS) Method with a random tree (RT) classifier, named RSRT, to create a reasonable set of base learners. Their proposed model can construct a set of random trees using different randomly selected subsets of features from all different random features of the training dataset. As a result, the ensembles of trees are able to reduce redundancy of features and prevent the system to overfit, which keeps the strength of the individual trees over the split random selection. Results show their model achieved high attack detection rates on ICS power system cyber attack Mississippi State University dataset~\cite{hink2014machine}.

\begin{figure*}
\centering
\includegraphics[width=0.72\textwidth]{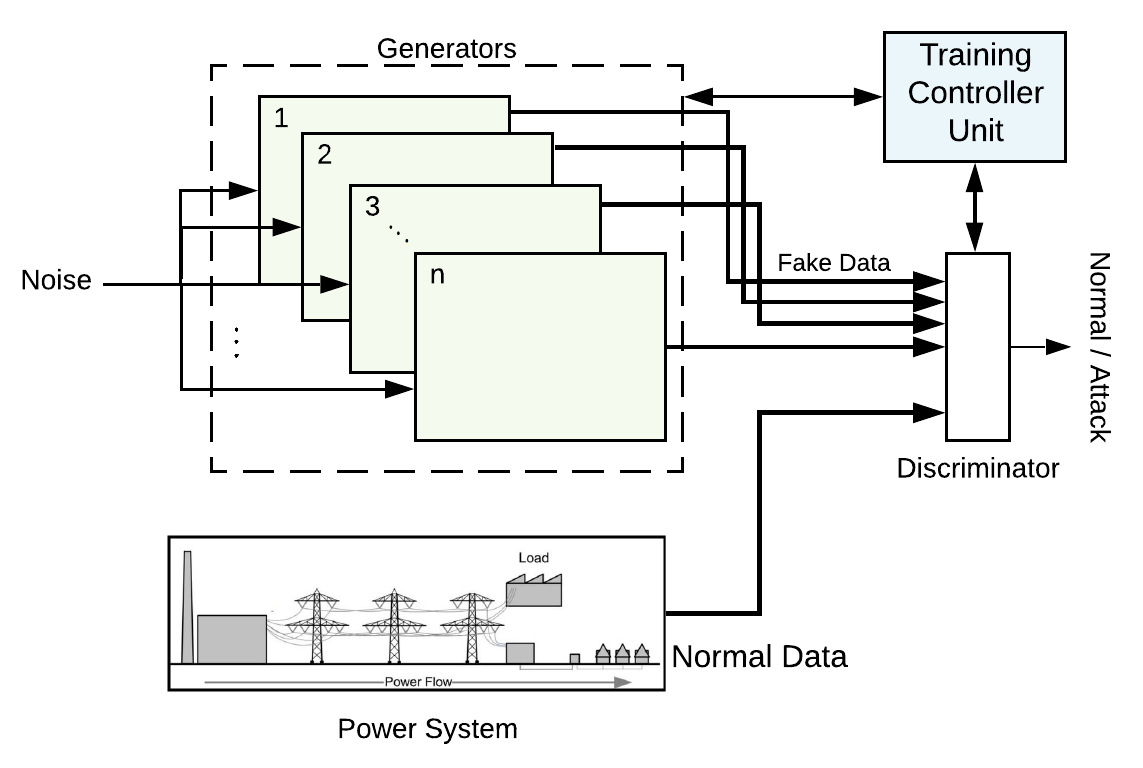}
\caption{Architecture of the Proposed System.}
\label{fig:architecture}
\end{figure*}

\section{Method}
\label{sec:proposed-system}
The main task of anomalies detection for sequential data is to determine whether the distributions of new observations fit the distribution of normal data learned in the training phase. Several terms are used to describe non-fitting points in different domains: anomalies, outliers, intrusions, failures, or contaminants.
We propose an innovative countermeasure that attempts to model and detect the distributions of possible types of attacks in an unsupervised fashion on power systems based on generative adversarial networks (GANs).
In the GAN objective, the task of a generator is usually more difficult than the task of the discriminator since the generator has to generate fake samples that maximize the mistake of the discriminator~\cite{ghosh2018multi}. This fact, in addition to the min-max nature of GAN objective function, leads to several issues for GAN-based models, such as \textit{mode collapse}, \textit{requiring large amounts of training data}, and \textit{difficult optimization}.
Unlike conventional GANs, the discriminator plays a more important role in our model. In addition, we tried to address the \textit{mode collapse} problem by increasing the number of generators as well as by providing an interaction between the discriminator and the generators, relying on the performance of the discriminator.
Our method is inspired by two GAN-based models, called one-class adversarial networks (OCAN)~\cite{zheng2019one} and multi-agent diverse generative adversarial networks (MAD-GAN)~\cite{ghosh2018multi}. The former was originally proposed for fraud detection on online applications such as social media. In the first step, OCAN captures the meaningful representation of normal users from their sequences of online activities using the long short term memory-autoencoder (LSTM-AE)~\cite{srivastava2015unsupervised}. In the next step, it trains a discriminator of a complementary GAN model in order to detect malicious users. The main difference between regular GAN and complementary GAN is that the generator in regular GAN learns to reconcile the distribution of the generated fake data representation to the representation of normal data. However, the generator in complementary GAN is trained to generate a distribution of data that is close to the complementary distribution of the normal data. The idea behind the complementary GAN is based on the fact that the distribution of attack data, which is capable of misleading systems, is similar to normal data but not exactly the same. As a result, it can intuitively be argued that the distribution of attack data is close to the complementary distribution of normal data. On the other hand, the MAD-GAN was originally proposed to resolve the well-known \textit{mode collapse} problem of regular GANs. To this end, the MAD-GAN introduces a multi-agent GAN framework that utilizes multiple generators and one discriminator. The generators are trained to generate diverse high probability regions of the real data distribution, resulting in different identifiable modes. The discriminator aims to differentiate between normal data and generated data by learning the distribution of generators and normal data. Whereas, the generators aim to maximize the discriminator error.

Our method uses $n$ generators as MAD-GAN, but tries to learn the complementary data distribution as OCAN. Constraints on the optimization ensure that the \textit{mode collapse} problem is reduced.

\subsection{Proposed System}
The overall architecture of the proposed system is shown in Figure. \ref{fig:architecture}. The model involves $n$ generators and one discriminator.
The discriminator $D(x;\theta_d)$ assigns each observation $x$ to one of $n+1$ classes determining if the observation has been generated by one of the $n$ generators or belongs to the real data (class $n+1)$.
To train our system, similarly to regular GANs, we apply prior input noise $z \sim p_z$ to the multiple generators.
Each generator $i$ outputs a fake sample $\tilde{x_i} = G_i(z;\theta_{g}^i)$ according to the distribution $p_{g_i}$.
The parameters $\theta_{g}^i$ of each generator are optimized by minimizing the objective function
\begin{equation}
    \mathop{\mathbb{E}}_{x \sim p_d} \log D_{n+1}(x:\theta_d) +\mathop{\mathbb{E}}_{z \sim p_z} \log (1-D_{n+1}(G_i(z;\theta_{g}^i);\theta_d)),
\end{equation}
where $D_{n+1}(x; \theta_d)$ is the probability that observation $x$ belongs to the real data, $p_d$ is the probability distribution of the real data, and $p_z$ is the distribution of noise.
The joint objective of all the generators is to minimize
\begin{equation}
    \mathop{\mathbb{E}}_{x \sim p_d} \log D_{n+1}(x) + \sum_{i=1}^n  \mathop{\mathbb{E}}_{x \sim p_{g_i}} \log (1-D_{n+1}(x)).
\end{equation}
Simultaneously, the objective of the discriminator, which is optimizing $\theta_d$, is  to maximize:
\begin{equation}
    \mathop{\mathbb{E}}_{x \sim p} H(\delta , D(x, \theta_d)), 
\label{eq:objective_discriminator}
\end{equation}
where $\delta \in \{0,1\}^{n+1}$, and for $i \in \{1,...,n\},
\delta(i) = 1$ if the sample belongs to the \textit{i}-th generator, otherwise $\delta(n + 1) = 1$ and $H(.,.)$ is the negative of the cross entropy function.
Obviously, the discriminator needs to learn to push different generators towards different identifiable modes in order to accurately recognize the generator that generated a given fake data. Nevertheless, the objective of each generator is the same as the objective of the generator in the standard GAN. Finally, in order to force the generators to generate data with wider distributions, we apply a condition on a min-max interaction between generators and discriminators. 
The generators keep learning as long as the sensitivities (SEs) and specificities (SPs)~\cite{lalkhen2008clinical} from the discriminator are above the values of two hyper-parameters $\alpha$ and $\beta$, respectively. When these values fall below the thresholds $\alpha$ or $\beta$, we pause training the generators until the discriminator learns to perform better than those thresholds values.

As a result, the proposed method objective can be written as
\begin{multline}
\label{eq:objective_function}
    \min_{\substack{\theta_g\\(\text{SE} > \alpha, \text{SP} > \beta)}}\max_{\theta_d} V(\theta_d,\theta_g) :=  \mathop{\mathbb{E}}_{x \sim p_d} \log D_{n+1}(x) + \\ + \sum_{i=1}^n  \mathop{\mathbb{E}}_{x \sim p_{g_i}} \log (1-D_{n+1}(x)) + \mathop{\mathbb{E}}_{x \sim p} H(\delta , D(x, \theta_d)).
\end{multline}

\begin{figure}
        \centering
        \begin{subfigure}[b]{0.235\textwidth}
            \centering
            \includegraphics[width=\textwidth]{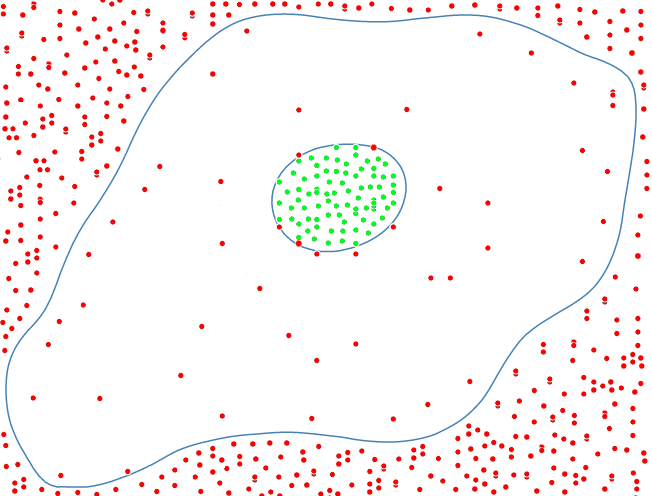}
            {{\small Epoch number = 1.}}    
        \end{subfigure}
        \hfill
        \begin{subfigure}[b]{0.235\textwidth}  
            \centering 
            \includegraphics[width=\textwidth]{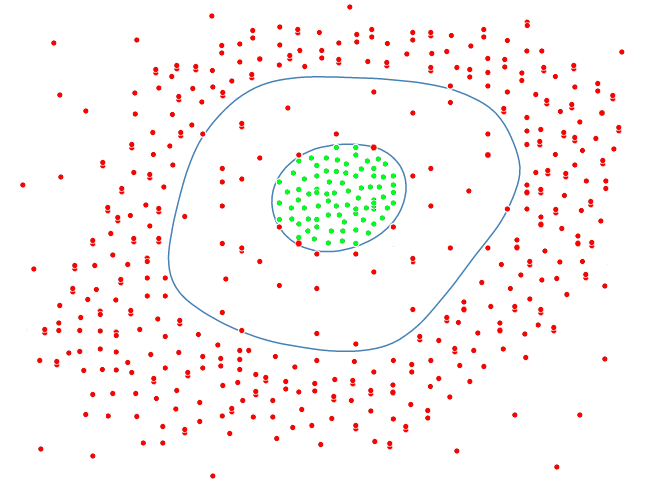}
            {{\small Epoch number =  10.}}
        \end{subfigure}
        \vskip\baselineskip
        \begin{subfigure}[b]{0.235\textwidth}   
            \centering 
            \includegraphics[width=\textwidth]{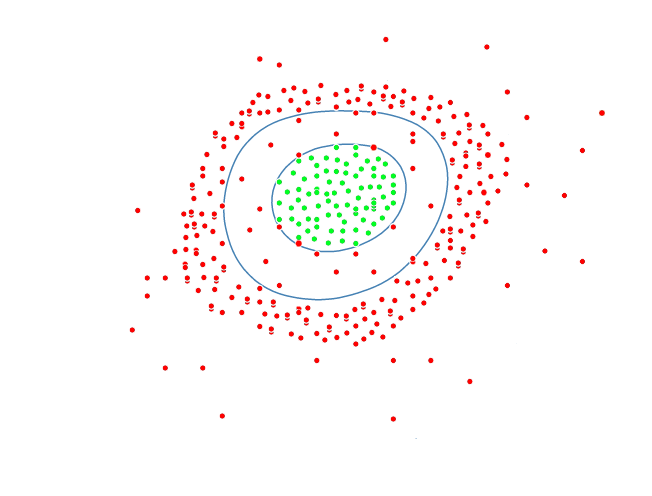}
            {{\small Epoch number =  50.}}
        \end{subfigure}
        \hfill
        \begin{subfigure}[b]{0.235\textwidth}   
            \centering 
            \includegraphics[width=\textwidth]{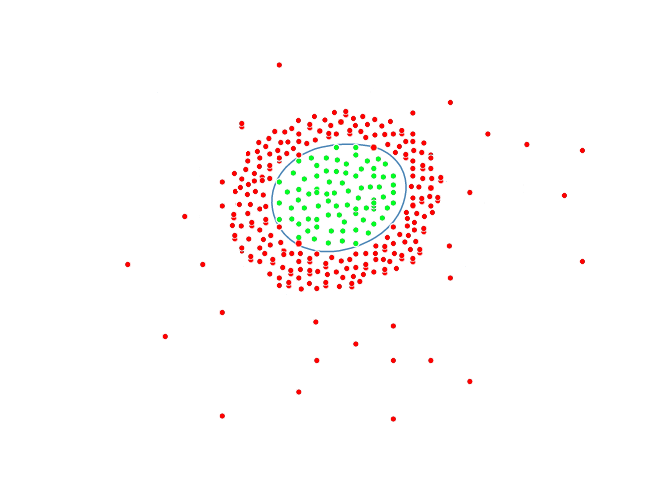}
            {{\small Epoch number = 100.}}
        \end{subfigure}
        \caption{The schematic of train cycle. Green points indicate normal data and red points represents generated data in each cycle. In each training cycle, the generated data converges to the complementary distribution of normal data.}  
        \label{fig:train_cycle}
    \end{figure}

The training process is illustrated in Figure.~\ref{fig:train_cycle}.
Similar to regular GAN, the objective functions of the generators try to learn the distribution of the normal data and then generate fake data that has a minimum difference with the distribution of normal data. However, the discriminator tries to learn to maximally discriminate the generated fake samples from real data. One of the main differences between our model and regular GAN is that the discriminator identifies the generated data that belongs to different generators (Eq.~\ref{eq:objective_discriminator}). It then maximizes the dissimilarity of the distributions corresponding to each generator. This is similar to MAD-GAN, but is not sufficient to resolve the \textit{mode collapse} problem completely (as we will illustrate later in Figure.~\ref{fig:method_comparision}).
We improve the resolution of mode collapse by employing two thresholds in the training phase as explained above. Inspired by this step-by-step training process, we named our proposed model \textit{STEP-GAN}. Using this strategy, we do not fully optimize the generators. This results in more spread distributions that can better cover the complementary distribution of normal data (see Figure.~\ref{fig:method_comparision}-c). This cycle repeats until it guarantees that the discriminator is well trained to detect anomalies from real data, and the generators are able to generate a wide distribution of possible attach data.

The algorithm of the proposed attack detection model is shown in Algorithm~\ref{al:algorithm}.
Both the objective function Eq.~\ref{eq:objective_function} and the interaction between the generators and the discriminator are intended to ensure that the generated fake samples are not only well distributed in the plane, but they are also similar to the real-world attack samples. If this is true, it can be argued that the generated fake data (simulated attack samples) by the generators can be sued as a simulation of potential real-world attacks to power systems.

\begin{algorithm}[!htb]
\setstretch{1.35}
\SetAlgoLined
\KwResult{Anomalies Detection in Power Systems.}
 \% Training phase\;
 \While{(Epoch number $<$ Max epoch)}{
    \While{(\textbf{SE} $< \alpha$) and (\textbf{SP} $< \beta$)}{
     \begin{spacing}{1.7}
     \end{spacing}
     Train Discriminator().\\
    }
    Train Generators().\\
 }
 \% Testing phase\;
 \begin{spacing}{1.6}
 \end{spacing}
 Discriminate normal data from attack data by $D_{n+1}(x)$.
\setstretch{1}

\begin{spacing}{1.8}
\end{spacing}
\caption{The proposed method algorithm. SE and SP indicate \textit{Sensitivity} and \textit{Specificity}, respectively. hyper-parameter $\alpha$ and $\beta$ are  adjustable thresholds.}
\label{al:algorithm}
\end{algorithm}


\subsection{Evaluation}
During the evaluation, the trained discriminator decides which class the input data belongs to (attack or normal data) using a softmax function. Being this a binary classification problem, we use accuracy and F-measure to evaluate the system performance. We also used a 10-fold cross-validation strategy for training and testing the proposed model. In this strategy, we randomly assigned 10\% of the dataset to the test set and the rest of the normal data are used to train the system. This division will be repeated ten times, and the testing result is the average of testing results for the ten times.

\section{Experimental setup}
\label{sec:Experimental-setup}

The generators are composed of three fully connected layers with 50, 300 and 128 nodes for each layer, respectively. All the hidden layers use Parametric ReLU (PReLU)~\cite{he2015delving} activation functions, whereas the last layer uses Tanh activation function. The discriminator consists of 6 fully connected layers (4 hidden layers). The input layer of the discriminator consists of 128 nodes. Each hidden layer includes 300 nodes with Leaky ReLU activation function~\cite{zhang2017dilated}, whereas the output layer has $n+1$ softmax nodes that compute the probability of an input sample to be a normal or attack. We use the Adam optimizer~\cite{kingma2014adam} for training generators and the discriminator. The cross-entropy is used as a loss function.

The main experimental parameters that are varied in our experiments are the number of generators in the model that can assume any value in $\{1, 2, 3, 5, 10, 15, 20\}$ and the hyper-parameters $\alpha$ and $\beta$ that are varied between $0.55$ and $1.0$ in intervals of $0.05$.
All models are implemented in PyTorch~\cite{paszke2017automatic}.

\subsection{Baseline systems}
In order to evaluate the performance of our method, we compare it to the two architectures the proposed method is inspired by, namely one-class adversarial networks (OCAN) and multi-agent diverse generative adversarial networks (MAD-GAN). Similar to regular GAN, the OCAN uses one generator and a complementary discriminator. However, MAD-GAN uses multiple generators and one discriminator. We also examined the MAD-GAN model with various number of generators as we do for the proposed model.

\begin{figure}
\centering
\includegraphics[width=0.48\textwidth]{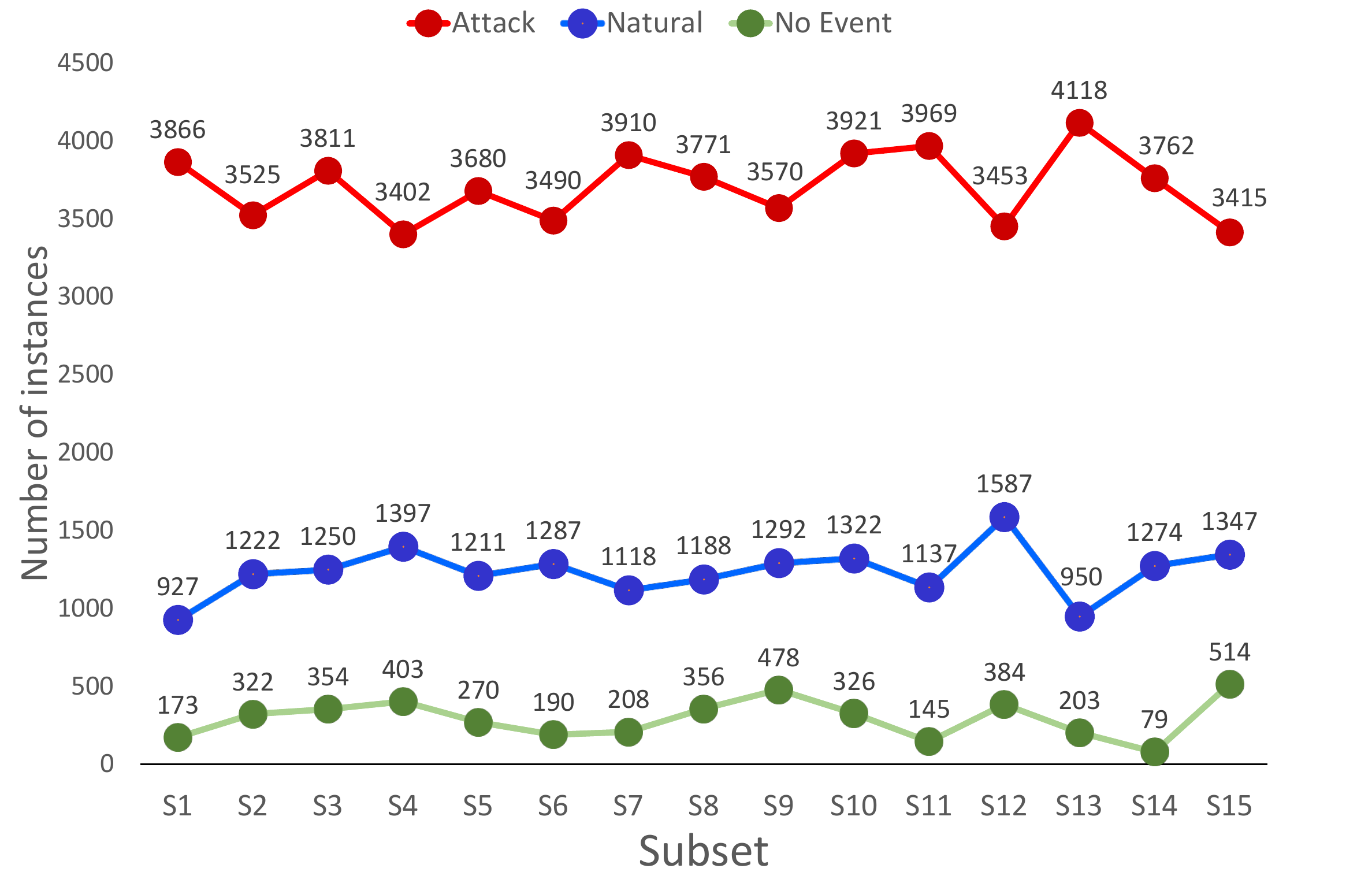}
        \caption{Distribution of benchmark datasets.}
        \label{fig:dataset}
\end{figure}

\subsection{Dataset}
The experiments are conducted on an open-source simulated ICS cyber attack dataset obtained from Supervisory Control and Data Acquisition (SCADA) power systems provided by Mississippi State University~\cite{hink2014machine}. The dataset contains three groups including Binary, Three-Class, and Multiclass datasets. Each group is made from one initial dataset including 15 subsets that consist of 37 power system event scenarios, compromising 28 \textit{Attack Events}, 1 \textit{No Events}, and 8 \textit{Normal Events}. The benchmark distribution of each dataset instances is shown in Figure.~\ref{fig:dataset}. Each instance includes 128 fixed-length dimensional sequential data. In this study, we used binary classification events for detecting the FDI attacks on the SCADA system. In order to reduce the effect of a small sample size, the datasets were randomly sampled at 1\% to reduce the size and evaluate the effectiveness of small sample sizes. The dataset statistics of binary class events classification is summarized in \ref{tbl:database}.

\begin{table}[!htb]\normalsize
\begin{center}
\caption{The dataset statistics of binary classification events.}
\label{tbl:database}
\begin{tabular}{|l|c|c|}
\hline
Subset        & \#Event Scenario & \#Instances \\ [0.5ex]
\hline\hline
No Events     & 1                & 294 \\
Natural Event & 8                & 1221 \\
Attack        & 28               & 3711 \\
\hline
Total         & 37               & 5226 \\
\hline
\end{tabular}
\end{center}
\end{table}

\subsection{Evaluation Metrics}
\label{sec:eval-met}
The detection of cyber-physical attacks in power systems is a binary classification task, in which data sequences from real sources (No Events or Natural Events) are labeled as positive classes and attack data are labeled as negative classes. Therefore, to verify the performance of the proposed method we used two metrics: Accuracy and F-measure defined as follows:
\begin{eqnarray}
        \text{Accuracy} &=& \frac{T_P + T_N}{T_P + T_N + F_P + F_N},\\
        \text{F-measure} &=& \frac{2 \times T_P}{2 \times T_P + F_N + F_P},
\end{eqnarray}
where $T_P$, $T_N$, $F_P$ and $F_N$ indicate \textit{true positive}, \textit{true negative}, \textit{false positive} and \textit{false negative}, respectively.

\begin{figure}
\centering
\includegraphics[width=0.45\textwidth]{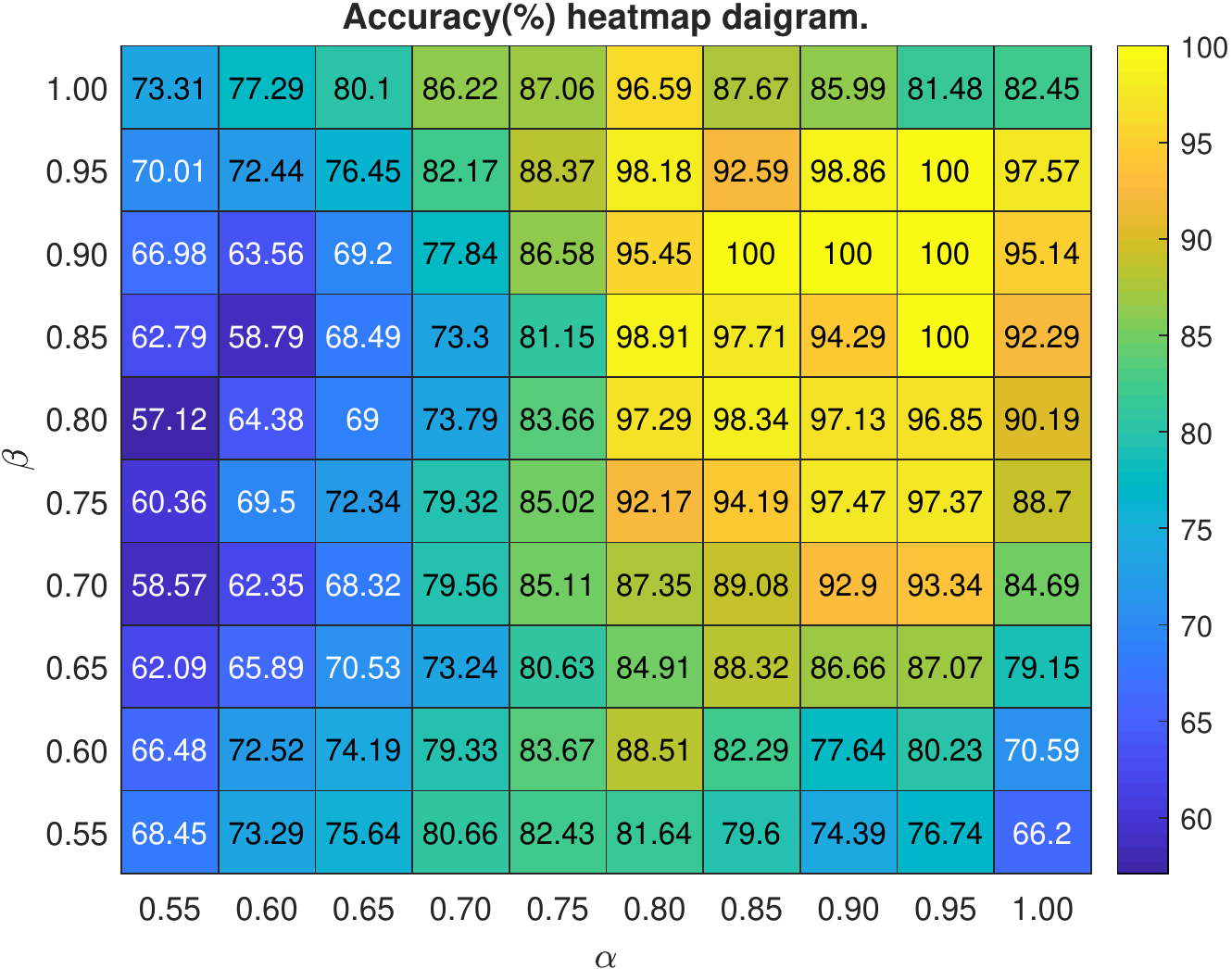}
        \caption{The heat-map diagram of the results obtained from the proposed model for different values of $\alpha$ and $\beta$ and a fixed ten generators.}
        \label{fig:heatmap}
\end{figure}

\begin{table}
\centering
\setlength\tabcolsep{1.5pt} 
\caption{Average results of the proposed method (STEP-GAN) in terms of accuracy(\%) on 15 subsets.}
\label{tbl:proposed-method-results}
\begin{tabular}{cccccc}
\hline\hline
 \#Generators & \multicolumn{5}{c}{Hyper-parameters ($\alpha$, $\beta$)} \\
 \hline
 &
  (0.95,0.95) &
  (0.9,0.9) &
  (0.8,0.8) &
  (0.7,0.7) &
  (0.6,0.6) \\
  \hline
  1 & 68.56 & 71.06 & 59.27 & 53.91 & 43.81 \\
  2 & 89.78 & 92.36 & 78.81 & 57.50 & 50.07 \\
  3 & 96.45 & 97.40 & 82.67 & 66.89 & 55.99 \\
  5 & \textbf{100.0} & \textbf{100.0} & 95.44 & 74.30 & 67.31 \\
  10 & \textbf{100.0} & \textbf{100.0} & 97.29 & 79.56 & 72.52 \\
  15 & 97.31 & \textbf{100.0} & 92.11 & 69.21 & 68.45 \\
  20 & 95.82 & \textbf{100.0} & 89.26 & 64.18 & 63.83 \\
  \hline\hline
\end{tabular}
\end{table}

\begin{figure*}
\centering
\includegraphics[width=1.0\textwidth]{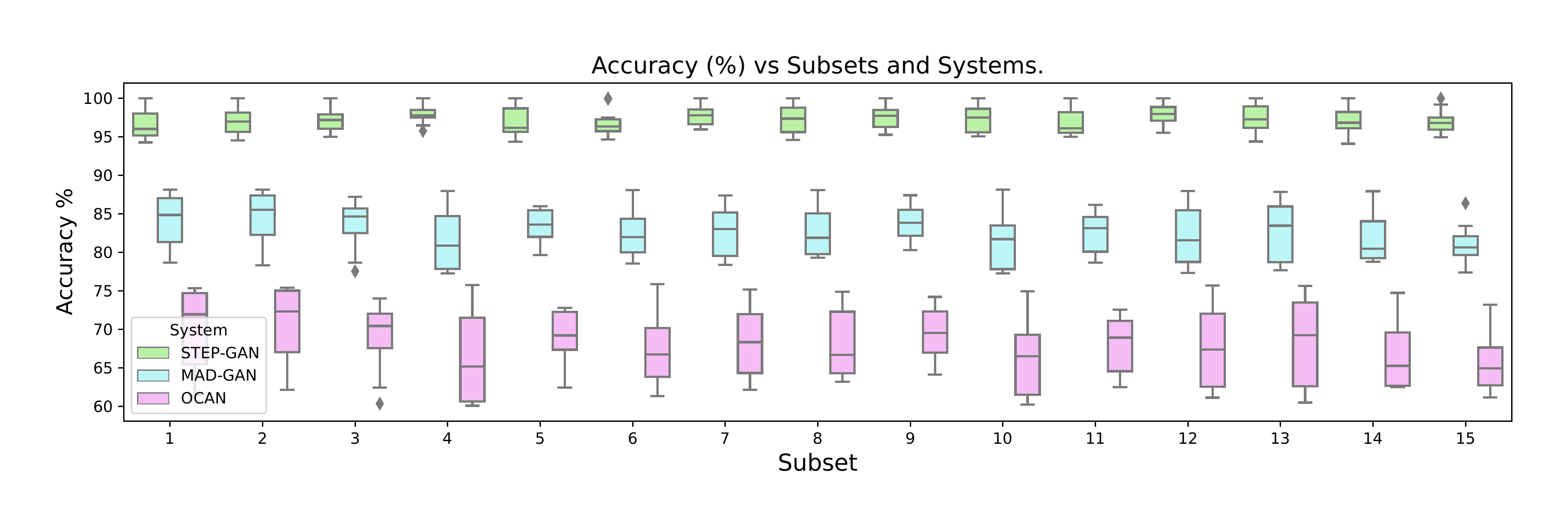}
        \caption{A comparison of STEP-GAN (top), MAD-GAN (middle) and OCAN (bottom) for each subset. The results are obtained from the best configurations of each system for 10 independent repetitions of training the systems. In the case of STEP-GAN, $\alpha$ and $\beta$ are set 0.9 and 10 generators are used for training the STEP-GAN.}
        \label{fig:comparision}
\end{figure*}

\section{Results}
\label{sec:results}
Table~\ref{tbl:proposed-method-results} shows average accuracy as a function of the number of generators and different values for the hyper-parameters. The performance is strongly dependent both on the number of generators and on the values of the hyper-parameters. Best results are obtained for more than 5 generators and for $\alpha$ and $\beta$ over 90\%.
In order to examine the effects of $\alpha$ and $\beta$ on the proposed model, we fixed the number of generators to 10 and varied $\alpha$ and $\beta$ with finer intervals. The corresponding results are show in Figure.~\ref{fig:heatmap}. The results show that variations in $\alpha$ values affect the performance of the model more than $\beta$ variations. This means that the model is more dependent on the performance of the discriminator for detecting normal data (higher value of sensitivity) than the performance of generators for simulating fake data (higher values of specificity).
In the following we will use our method with hyper-parameters $(\alpha, \beta) = (0.9,0.9)$.

\begin{table}
\centering
\caption  {A comparison of the best results obtained from the different numbers of the proposed system (STEP-GAN) generators and baseline systems.}
\label{tbl:results_comparision}
\begin{tabular}{lccccc}
\hline\hline
System & \#Generators & Accuracy\% & F-measure  \\ [0.5ex]
\hline
RSRT~\cite{hassan2020increasing} & -             & \textbf{95.95}        & not reported          \\
\hline
OCAN                             & (Default = 1) & \textbf{70.64}        & \textbf{0.3510}          \\
\hline
                             & 1            &  64.77       &  0.3021       \\ 
                             & 2            &  69.46       &  0.3605       \\
                             & 3            &  74.32       &  0.5984       \\
MAD-GAN                      & 5            &  77.20       &  0.6237       \\ 
                             & 10           & 80.91 & 0.6629          \\
                             & 15           & 83.05        & 0.7044         \\
                             & 20           & \textbf{87.53} & \textbf{0.7390}          \\
\hline
                             & 1            & 71.06      &  0.6789       \\ 
                             & 2            & 92.36      &  0.9476       \\
                             & 3            & 97.04      & 0.9601          \\
Step-GAN                     & 5            & \textbf{100.00}        & \textbf{1.0000}          \\
(proposed)                   & 10           & \textbf{100.00}        & \textbf{1.0000}          \\
                             & 15           & \textbf{100.00}        & \textbf{1.0000}          \\
                             & 20           & \textbf{100.00}        & \textbf{1.0000}          \\
\hline\hline
\end{tabular}
\end{table}

Table~\ref{tbl:results_comparision} shows a comparison between the proposed model and the baseline models. We also included results reported in \cite{hassan2020increasing} for comparison.
From the table, it is clear that STEP-GAN outperforms MAD-GAN and OCAN for this task in all configurations, that is, both in absolute terms but also for an equal number of generators. STEP-GAN with 5 or more generators also outperforms RSRT~\cite{hassan2020increasing}.
The best configuration for our model and the baseline models are compared in Figure~\ref{fig:comparision}.
The boxplots show the results over 10 independent repetitions, and for each subset of the database.
This figure shows how the performance improvement for our system is consistent and stable over repetitions and over subsets of the database.

\begin{figure*}
     \centering
     \begin{subfigure}[b]{0.33\textwidth}
         \centering
         \includegraphics[width=\textwidth]{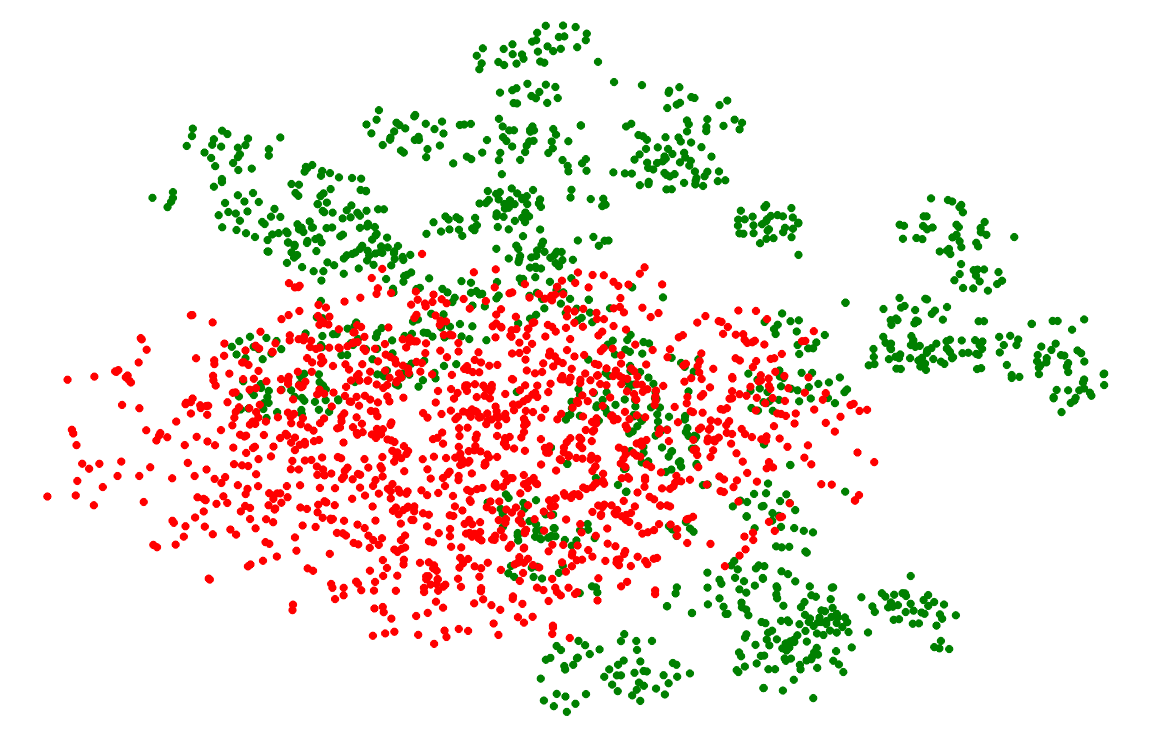}
         \caption{OCAN}
         \label{fig:OCAN}
     \end{subfigure}
     \hfill
     \begin{subfigure}[b]{0.33\textwidth}
         \centering
         \includegraphics[width=\textwidth]{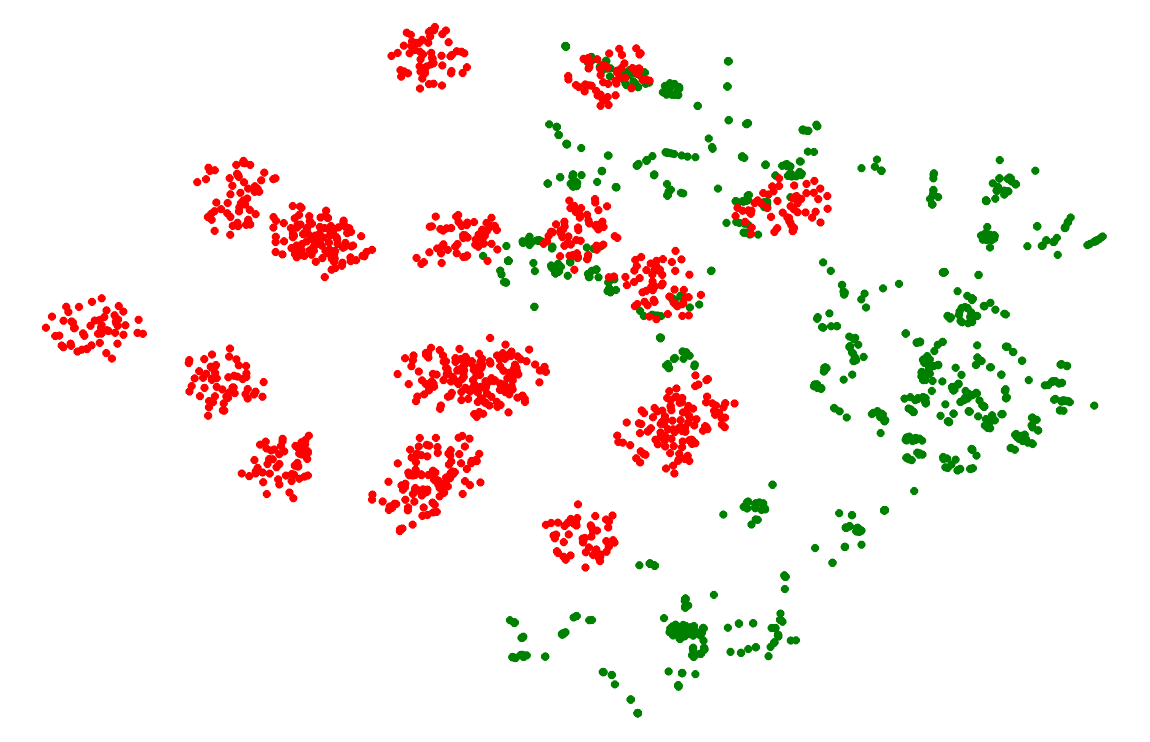}
         \caption{MAD-GAN}
         \label{fig:MAD-GAN}
     \end{subfigure}
     \hfill
     \begin{subfigure}[b]{0.33\textwidth}
         \centering
         \includegraphics[width=\textwidth]{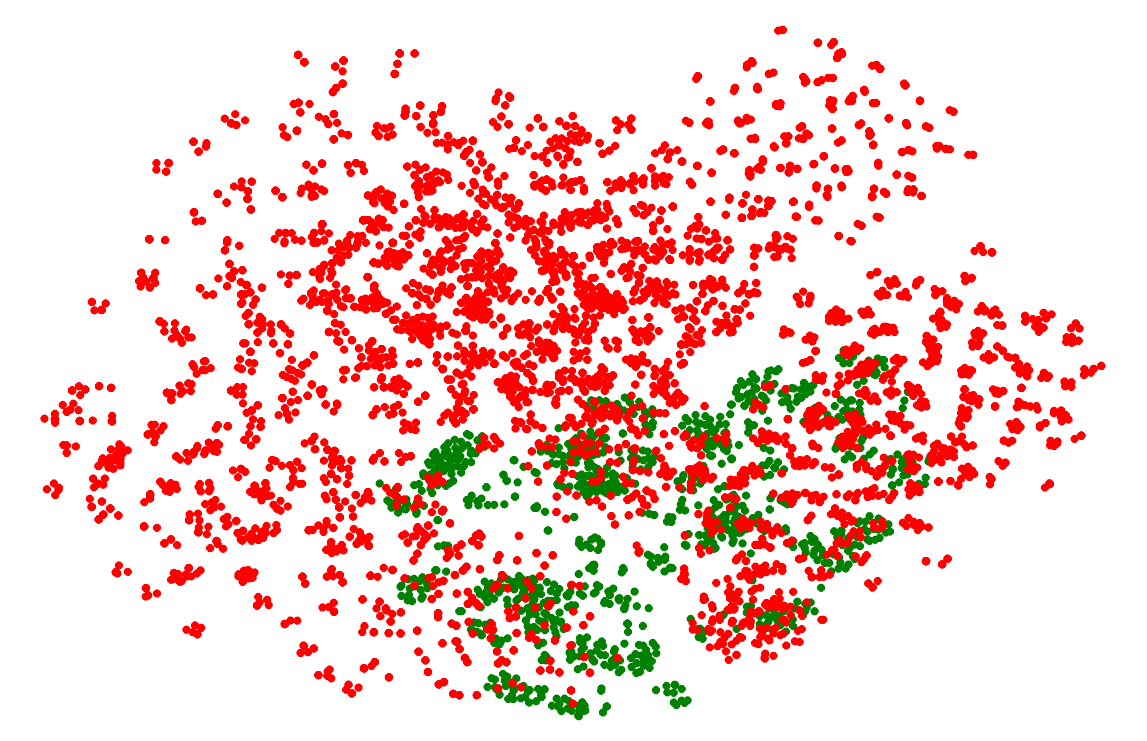}
         \caption{STEP-GAN}
         \label{fig:STEP-GAN}
     \end{subfigure}
        \caption{A t-SNE comparison of generated data by (a) OCAN, (b) MAD-GAN, and the (c) proposed method. Green points indicate normal data, and red points are generated by the model. As it can be seen, OCAN and MAD-GAN suffer from \textit{"mode collapse"}. While STEP-GAN covers the complementary distribution of normal data well.}
        \label{fig:method_comparision}
\end{figure*}

Figure~\ref{fig:method_comparision} illustrates the behaviour of the proposed systems when addressing the model collapse issue. We used t-SNE to project the original and generated data onto a two-dimensional space for visualization.
Compared to OCAN (Figure~\ref{fig:method_comparision}a) and MAD-GAN (Figure~\ref{fig:method_comparision}b), STEP-GAN (Figure~\ref{fig:method_comparision}c) obtains a wider coverage of the space surrounding the real data, and thus a reduced mode collapse problem with a lower number of generators.
As a consequence we obtain a more general model and prevent overfitting.

\begin{figure}
\centering
\includegraphics[width=0.43\textwidth]{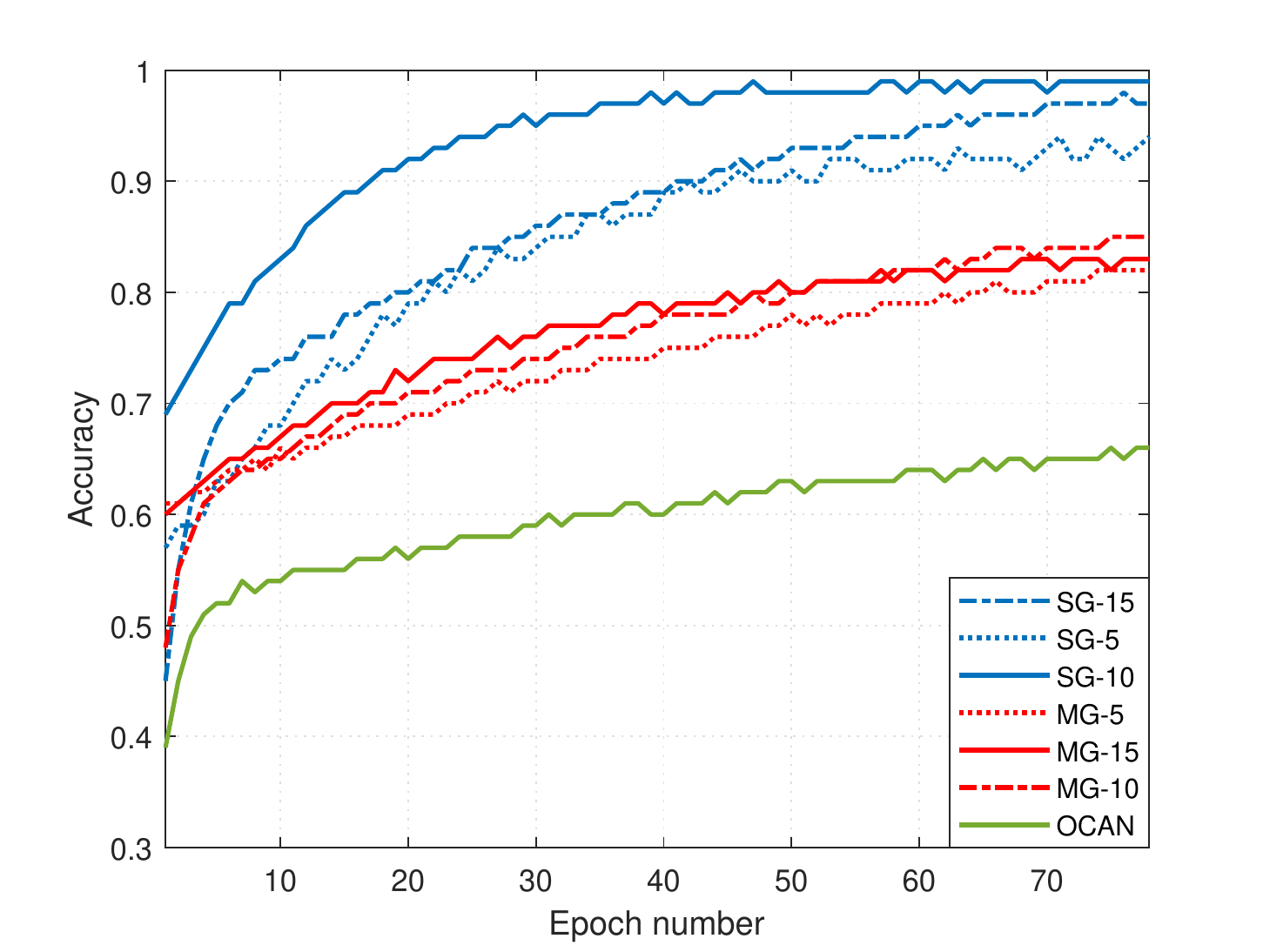}
        \caption{Epoch comparison of the proposed method (STEP-GAN), MAD-GAN, and OCAN. SG and MG represent "STEP-GAN" and "MAD-GAN", respectively, and the numbers in front of them indicate the number of generators.}
        \label{fig:epoch_comparision}
\end{figure}

Finally, Figure~\ref{fig:epoch_comparision} shows the convergence behavior of the proposed model compared to MAD-GAN and OCAN for the best of their configurations. As can be seen in Figure~\ref{fig:epoch_comparision}, STEP-GAN converges using fewer number of epochs. However, MAD-GAN and OCAN require a higher number of epochs to converge. The remarkable point is that the OCAN system, despite using only one generator, needs more epochs to converge, indicating the advantage of our proposed system in computational complexity.

\section{Discussion}
\label{sec:discussion}
A system that is successful in detection of cyber-attacks on power grids or other systems, should be able to predict data that has not been seen during training.
A GAN based system is particularly suitable for this kind of application since the generators can be encouraged to generate data outside the normal data distribution by choosing the proper optimality criterion.
When the distributions are complex, it is beneficial to employ several generators in order to explore the space more thoroughly.
However, doing this creates a trade-off: that can be seen in Figure~\ref{fig:method_comparision}: fewer generators may result in poor exploration of the space (Figure~\ref{fig:method_comparision}a), but more generators can result in overfitting and the phenomenon known as mode collapse (Figure~\ref{fig:method_comparision}b).
Our system can take advantage of the increased number of generators, and at the same time reduce mode collapse by carefully limiting the optimization of the generators during training, and therefore avoiding overfitting (Figure~\ref{fig:method_comparision}c).

On one hand, a high sensitivity value is crucial when only real data is available. Given the role assigned to the discriminator in our task, it is exceedingly important for the proposed system to be able to correctly detect normal data. This means that we expect the model to reach a high sensitivity ratio. A proper choice of $\alpha$ allows the discriminator to competently detect the normal data. Experimental results show that 0.9 is an optimal choice for $\alpha$ both in terms of accuracy and acceleration of the training process. On the other hand, we seek generators that can simulate the distribution of hypothetical attacks data, which logically fits into the complementary distribution of normal data. A good choice of the $\beta$ causes fake data to be generated in the complementary distribution of the normal data.
If the value of hyper-parameter $\beta$ is too high, the generators lose their freedom of action, and may not be properly trained. If the $\beta$ value is too low, the generated data might be similar to normal data. Experimental results show that in an interaction between the generators and the discriminator, the best results are obtained when the $\beta$ value is 0.9. Using the aforementioned step-by-step training process, the generators are able to gradually bring the generated data closer to the complementary distribution of the normal data. Consequently, the generators are able to simulate potential attack data and the discriminator is able to competently detect the normal data.

\section{Conclusion}
\label{sec:conclusion}
In this study, we proposed a novel unsupervised countermeasure against cyber physical false data injection attacks on power systems. Our model is capable of potentially being employed in a wide range of one-class classification tasks. A multi-generators GAN-based model was used through a step-by-step training with the interaction between generators and a discriminator in order to simulate possible attacks to the system. Results show that our model significantly outperforms baseline systems and results reported in the literature on the highly imbalanced publicly available ICS cyber attack power system dataset. Compared to previously proposed methods, the system achieves better performance with a lower number of parameters (fewer generators).
The reason for this performance improvement is that the proposed training procedure mitigates the mode collapse issue in GAN based systems, and therefore generates more "space-filling" distributions with fewer generators.

\bibliographystyle{IEEEbib}
\bibliography{strings,refs}

\end{document}